\documentclass[conference]{IEEEtran}
\usepackage{amsmath,amsfonts}
\usepackage{algorithmic}
\usepackage{array}
\usepackage{orcidlink}
\usepackage[caption=false,font=normalsize,labelfont=sf,textfont=sf]{subfig}
\usepackage{textcomp}
\usepackage[utf8]{inputenc}
\DeclareUnicodeCharacter{03C1}{\rho}
\usepackage{stfloats}
\usepackage{url}
\usepackage{longtable}
\usepackage{verbatim}
\usepackage{graphicx}
\def\BibTeX{{\rm B\kern-.05em{\sc i\kern-.025em b}\kern-.08em
    T\kern-.1667em\lower.7ex\hbox{E}\kern-.125emX}}
\usepackage{balance}
\begin{document}
\title{A Bayesian Optimization-Based AutoML Framework for Non-Intrusive Load Monitoring}
\author{\IEEEauthorblockN{Nazanin Siavash\,\hypersetup{pdfborder={0 0 0}}\orcidlink{0009-0000-4177-0632}}
\IEEEauthorblockA{\textit{Department of Computer Science} \\
\textit{University of Colorado Colorado Springs (UCCS)}\\
Colorado Springs, United States \\
nsiavash@uccs.edu}
\and
\IEEEauthorblockN{Armin Moin\,\hypersetup{pdfborder={0 0 0}}\orcidlink{0000-0002-8484-7836}}
\IEEEauthorblockA{\textit{Department of Computer Science} \\
\textit{University of Colorado Colorado Springs (UCCS)}\\
Colorado Springs, United States \\
amoin@uccs.edu}
}

\maketitle

\begin{abstract}
Non-Intrusive Load Monitoring (NILM), commonly known as energy disaggregation, aims to estimate the power consumption of individual appliances by analyzing a home's total electricity usage. This method provides a cost-effective alternative to installing dedicated smart meters for each appliance. In this paper, we introduce a novel framework that incorporates Automated Machine Learning (AutoML) into the NILM domain, utilizing Bayesian Optimization for automated model selection and hyperparameter tuning. This framework empowers domain practitioners to effectively apply machine learning techniques without requiring advanced expertise in data science or machine learning. To support further research and industry adoption, we present AutoML4NILM, a flexible and extensible open-source toolkit designed to streamline the deployment of AutoML solutions for energy disaggregation. Currently, this framework supports 11 algorithms, each with different hyperparameters; however, its flexible design allows for the extension of both the algorithms and their hyperparameters.
\end{abstract}

\begin{IEEEkeywords}
energy disaggregation, non-intrusive load monitoring, nilm, automated machine learning, bayesian optimization
\end{IEEEkeywords}

\section{Introduction} \label{intro}
\IEEEPARstart{T}{he} transition towards cleaner and more energy-efficient power generation methods is currently underway, marking a significant shift away from traditional fossil-based approaches. In the evolving landscape of the electric power industry, it is crucial to promote a balanced relationship between energy consumption and power supply. Achieving this balance requires the scientific allocation and flexible scheduling of demand-side resources aided by advanced information technology \cite{Han+2023}. Demand Side Management (DSM) has emerged as a vital component of the rapidly advancing smart grid. Its importance is underscored by the increasing integration of intermittent and variable renewable energy sources. The unpredictable nature of these sources presents a major challenge in maintaining real-time equilibrium between demand and electricity production, especially in the absence of costly reserve services. DSM addresses this challenge by adjusting demand through shifts or reductions in consumption, enabling efficient use of available generation while ensuring minimum reserve capacity. One appealing strategy for Demand Side Management (DSM) is Direct Load Control (DLC), which allows utilities to influence customers' energy consumption patterns by remotely managing a predetermined set of controllable appliances \cite{Welikala+2017}. While a smart meter installed at the customer's location can enhance the flexibility of these appliances' loads, it is essential for the grid operator to have knowledge of the available flexible load at any given time. Without this information, utilities must maintain significant reserves, relying on generator reloads to meet demand.
Enhancing demand-side energy efficiency through effective energy monitoring is a crucial aspect of optimizing power consumption. Energy monitoring provides a detailed understanding of appliance energy usage and is fundamental to achieving intelligent power consumption services \cite{He+2019}. Among the methods used for load monitoring, Intrusive Load Monitoring (ILM) stands out as a conventional approach. ILM involves installing data acquisition devices for each load, offering accurate and reliable information \cite{Yin+2023}. However, the privacy concerns associated with accessing data from electricity usage and demand meters, along with the high implementation costs and complexity, render ILM impractical. In contrast, Non-Intrusive Load Monitoring (NILM) has gained widespread usage for acquiring appliance-level data due to its cost-effectiveness \cite{Cimen+2020}.

The concept of NILM was initially proposed by Hart at MIT and has since garnered significant attention from researchers \cite{Hart1992}. In contrast to ILM, NILM does not necessitate physical access to customers' premises. Instead, it employs a monitoring device positioned at the customer's electricity entry point or utilizes existing smart meters to collect bus data \cite{SchirmerMporas2023}. This data is then subjected to a series of algorithms that analyze and extract information regarding the energy consumption of specific appliances within the customer's premises. By obviating the requirement for appliance-specific sensors, NILM presents notable benefits in terms of reduced economic and labor costs associated with installation and maintenance. There are two main approaches for NILM analysis: event-based and state-based methodologies. Event-based techniques focus on detecting and categorizing events, which represent changes in aggregated energy consumption. This methodology involves three primary stages: event detection, feature extraction, and classification \cite{HeKanghang+2019}. Event detection aims to identify alterations in the aggregate signal that occur when one or more appliances transition their states. Feature extraction involves isolating electrical attributes for each detected event \cite{Kotsilitis+2023}. The subsequent classification and pattern matching stage are utilized to assign events to predefined categories corresponding to specific appliances. Event-based methods have a drawback in that they can sometimes lead to underestimation or overestimation of the load \cite{Kumar+2023}. This is because the discrepancy between the aggregated load and the sum of individual appliance power consumption is not typically taken into account during the classification step. On the contrary, state-based techniques allocate the overall power status of the system to a combination of working states of appliances \cite{Kumar+2023}, diverging from the analysis of individual events. These methodologies aim to identify the new combined state that the system has reached by modeling each appliance's operation as a state machine with distinct state transitions based on usage patterns \cite{Azizi+2021}. Consequently, state-based techniques demonstrate particular effectiveness in efficiently addressing the NILM problem for load signals with low sampling rates. Although state-based approaches often involve higher computational costs compared to event-based methods, they tend to yield greater accuracy.

In this study, we introduce an innovative method utilizing Bayesian Optimization to facilitate Automated Machine Learning (AutoML) for NILM. This method automates the selection of the optimal machine learning model and algorithm, tunes hyperparameters, and identifies a suitable configuration for training. As a result, it enhances the convenience and accessibility of state-of-the-art NILM techniques for practical deployment. Furthermore, it demonstrates significantly improved efficiency and superior experimental performance compared to traditional manual approaches, which typically evaluate only one model at a time. This work builds upon prior work by Moin et al. \cite{Moin+2025}  by extending its AutoML framework with three additional deep learning–based disaggregation algorithms, namely Window-based GRU, Seq2Seq, and Seq2Point, and by exploring a different hyperparameter search space to further improve generalizability. 

The structure of this paper is as follows: Section \ref{background} provides a background on NILM; Section \ref{literature-review} presents a literature review of NILM based on ML approaches. In Section \ref{method}, the AutoML algorithm is explained in detail. The experimental setup and results are shown in Sections \ref{experimental-setup} and \ref{experimental-results}, and finally, the conclusion is drawn in Section \ref{Conclusion}.

\section{Background} \label{background}
The objective of NILM is to determine the electrical loads of individual devices that contribute to the overall data recorded by the meter. This involves analyzing an aggregate power signal, denoted as  \(P\), which consists of a series of \(T\) samples measured over a specific period \(P={ρ(1),ρ(2),…,ρ(T)}\). Taking into account the presence of \(K\)appliances, the disaggregation process involves inferring the distinct power signatures  \(P\) associated with each appliance that constitutes the combined power signal. Consequently, the aggregate reading of the meter at a given time sample \(t\) can be mathematically represented as \cite{Nalmpantis+2019}:

\[\rho(t)= \sum_{k=1}^{K} p_k^{(t)} + \epsilon^{(t)}\]

\begin{figure*}
\begin{center}
    \centering
    \includegraphics[width=1\linewidth]{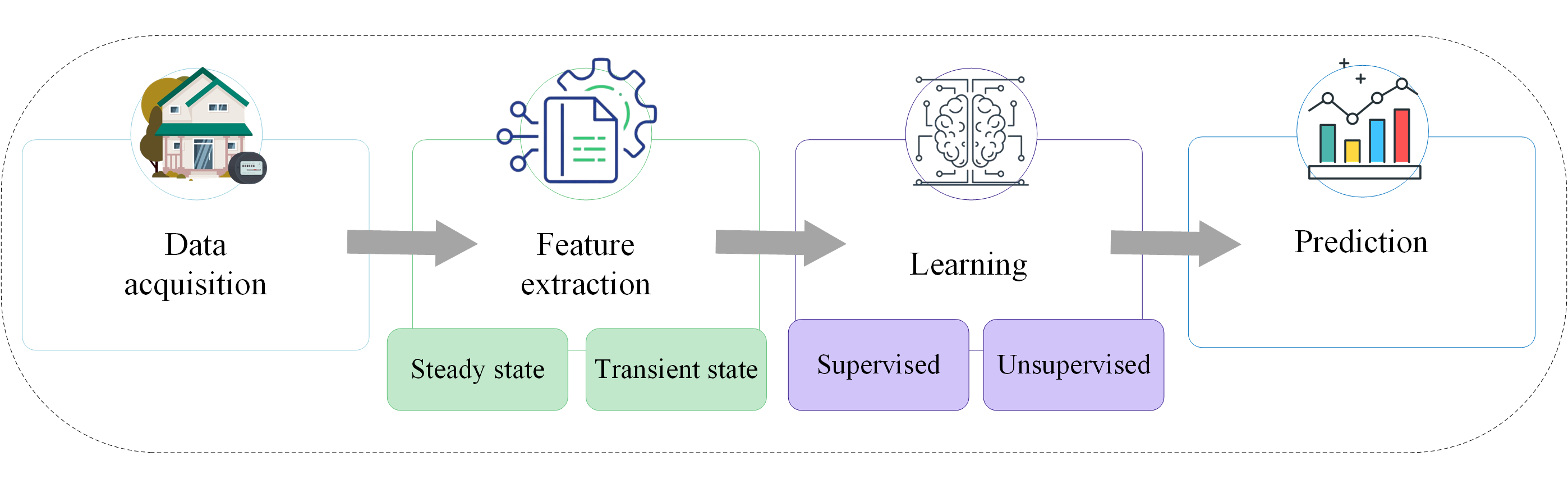}
    \caption{Framework Overview of NILM}
    \label{nilm-process}
    \end{center}
\end{figure*}

Where \(p_k^{(t)}\) is the power reading of the appliance \(k\) at time \(t\) and \(\epsilon^{(t)}\) represents model noise. 
The NILM analysis contains several stages, as depicted in Figure \ref{nilm-process}. Data acquisition is the initial step where raw data is collected based on specific criteria. The subsequent stage involves the extraction of load features during both steady and transient states. NILM utilizes these extracted features to identify load signature patterns for different appliances. Finally, the extracted features are employed for training and categorization purposes to identify and distinguish between various loads. The features can be categorized into steady state and transient state characteristics \cite{Sreevidhya+2019}. Steady state features encompass parameters such as active and reactive power, current harmonics, and V-I trajectory, as documented in existing literature. Conversely, transient states encompass factors like transient noise and inrush current.
The next stage is learning, which can be classified as either supervised or unsupervised. These concepts will be briefly reviewed in the following.

\textit{1) Supervised Learning (SL): }SL, also known as supervised ML, necessitates the labeling of data, prior to conducting training. This approach is employed when working with labeled energy datasets. Despite its notable performance, the requirement for labeled data poses challenges in real-world scenarios \cite{Jacobs+2020}. SL encompasses four primary subsections: classification models, deep neural networks (DNNs), statistical models, and regression \cite{Himeur+2023}. 

\textit{2) Unsupervised Learning (USL): }On the other hand, in unsupervised machine learning techniques, there is no requirement for pre-labeling prior to commencing the training process. However, it is important to note that unsupervised learning algorithms tend to be more costly to implement compared to supervised techniques. As a result, supervised learning techniques are commonly preferred for smaller applications. USL can be further divided into two main subsections: clustering and dimensionality reduction \cite{Himeur+2023}. Finally, after the learning process, the disaggregated energy will be obtained.
\section{Literature review} \label{literature-review}
In recent years, NILM methods have gained significant attention, particularly through the application of machine learning techniques. These approaches offer promising advancements in disaggregating electrical consumption data, enhancing energy efficiency and consumer insights. For example, Hidden Markov Models (HMM) and their variants, such as Factorial Hidden Markov Models (FHMM) \cite{ZHANGZhaoxia+2020}, are commonly employed as Unsupervised Learning (USL) approaches for constructing Finite State Machines (FSMs) due to their ability to capture probabilistic transitions between states. In addition to HMMs, other USL approaches have been investigated in the literature, including the iterative k-means clustering algorithm \cite{Sahrane+2021}, the Iterative Fuzzy C-Means clustering algorithm (IFCM) \cite{Ji+2019}, and the g-means clustering algorithm \cite{Maniar+2019}. Conversely, there has been a surge in the publication of household electricity datasets, which include both mains electricity readings and appliance power readings, enabling the formulation of NILM as a Supervised Learning (SL) problem. Deep neural networks offer the advantage of automated feature extraction, leading to reduced time consumption and improved performance in NILM \cite{Steephen+2022}. Recent literature has introduced several neural network architectures, including Convolutional Neural Networks (CNN) \cite{Shin+2019,Teixeira+2021,Chen+2021,Kong+2020}, tree-CNN \cite{Jia+2019}, Support Vector Machines (SVM) \cite{Wang+2018}, Recurrent Neural Networks (RNN) \cite{Chen+2021,Sayilar+2023}, Generative Adversarial Networks (GAN) \cite{Han+2023}, Naïve Bayes \cite{Yang+2018}, K-Nearest Neighbors (KNN) \cite{Yang+2018, Murtadha+2023}, and Long Short-Term Memory (LSTM) networks \cite{Heo+2021}. 
To streamline the process of selecting models and optimizing parameters in traditional machine learning methods, recent research has focused on NILM using AutoML approaches \cite{Meng+2023,Meng+2023AutoML}. In reference \cite{Meng+2023}, the authors compared three different open-source AutoML frameworks—Auto-Keras, AutoGluon-Tabular, and H2O—in the context of NILM, where active power consumption serves as the input feature. The experimental findings indicated that H2O and AutoGluon outperform Auto-Keras in terms of prediction performance. In reference \cite{Meng+2023AutoML}, the researchers first decomposed the high-frequency power signal using empirical mode decomposition (EMD) technology. They subsequently employed the Auto-Keras and H2O frameworks for load disaggregation, thereby addressing the challenges associated with model creation and hyperparameter tuning. Although both of these research papers focus on AutoML, neither implements the framework from scratch.

\section{Method} \label{method}
\subsection{AutoML for NILM}

\begin{figure*}
\begin{center}
    \centering
    \includegraphics[width=2\columnwidth]{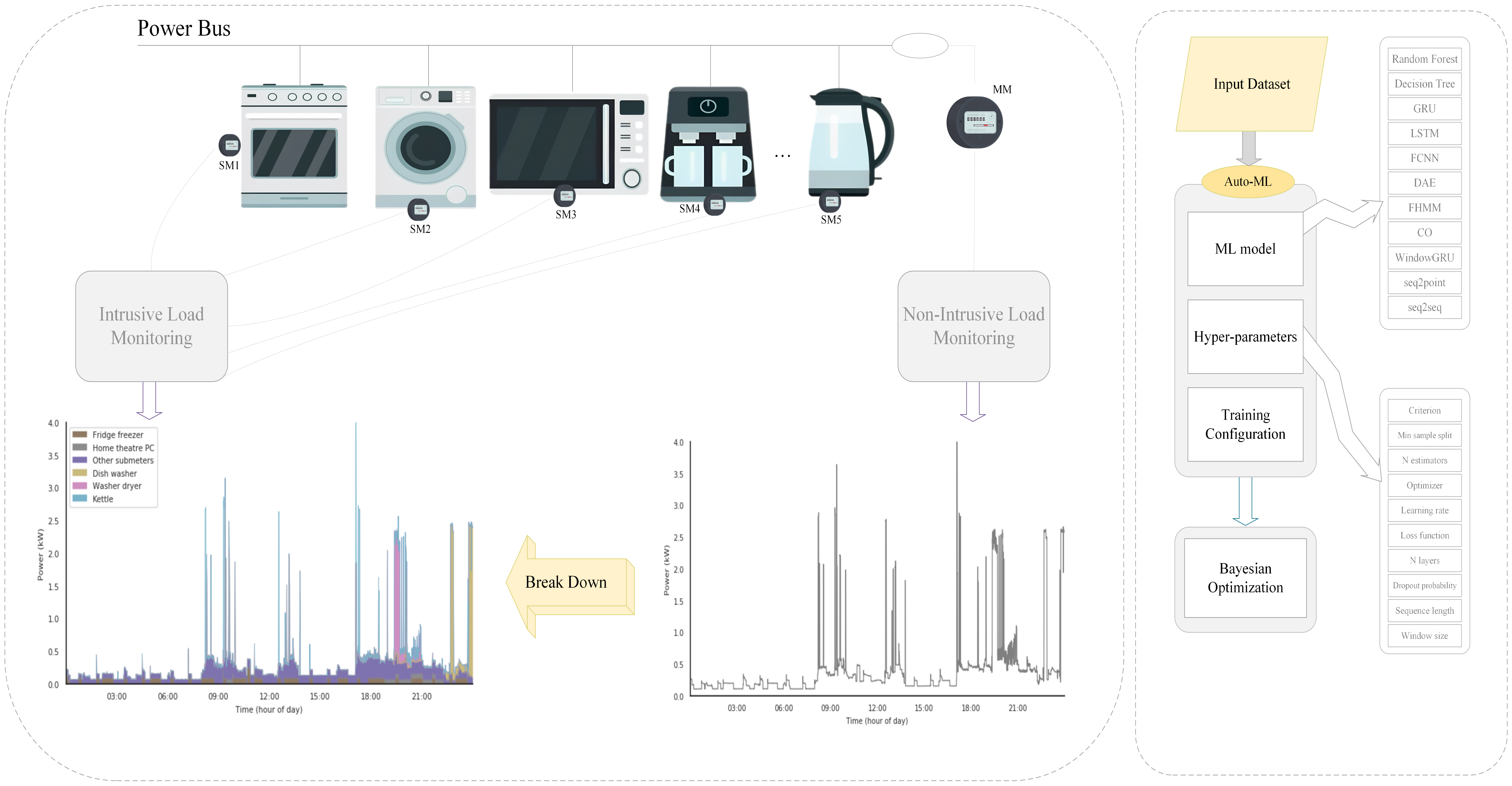}
    \caption{A comprehensive view of the presented work}
    \label{comprehensive-fig}
    \end{center}
\end{figure*}

AutoML aims to automate the design and development of machine learning systems, replacing manual trial-and-error processes with systematic, data-driven decision-making techniques \cite{Park+2020}. This automation broadens access to powerful ML tools, enabling a wider range of practitioners and domain experts to leverage advanced methodologies. In this paper, we propose a novel AutoML-based approach using Bayesian Optimization for the task of energy disaggregation.
Given a dataset \(D\), solving an ML problem typically involves defining a sequence of operators \(M = (O_1, \dots, O_z)\), where each operator \(O_i\) corresponds to a set of candidate ML algorithms. The associated hyperparameter spaces for these algorithms are represented by the set
$\Lambda = \left\{\left\{\Lambda_1^1, \dots, \Lambda_1^{n_1}\right\}, \dots, \left\{\Lambda_z^1, \dots, \Lambda_z^{n_z}\right\}\right\}.
$
The objective of the AutoML process is to discover the optimal Machine Learning  pipeline configuration, denoted by \(p(A_{1,\lambda}, \dots, A_{z,\lambda})^*\), that maximizes a given performance metric \cite{Liu2023}:
\[
p(A_{1,\lambda}, \dots, A_{z,\lambda})^* = \arg_{\lambda} \max f\left(p(A_{1,\lambda}, \dots, A_{z,\lambda})^l, D\right)
\]
where \(A_i\) denotes a specific algorithm chosen from operator \(O_i\), and \(\lambda\) refers to the set of hyperparameters. The function \(f\) measures the performance of a pipeline on the dataset \(D\).
To address the NILM task, we construct a comprehensive AutoML search space by evaluating various Machine Learning algorithms and their corresponding hyperparameters. Our proposed search space model is illustrated in Figure~\ref{comprehensive-fig}. Each Machine Learning model is explored with respect to its hyperparameters, including:
\begin{itemize}
    \item Criterion: \{squared error, friedman mse\}
    \item Minimum samples split: Uniform[10, 20]
    \item Number of estimators: Uniform[10, 30]
    \item Optimizer: \{Adam, Nadam\}
    \item Learning rate: \{1e-3, 1e-4\}
    \item Loss function: \{mean squared error, mean absolute error\}
    \item Number of layers: Uniform[5, 7, 1]
    \item Dropout probability: Uniform[0.1, 0.3]
    \item Sequence length: \{10, 20, 50\}
    \item Window size: \{20, 50, 100\}
\end{itemize}

Bayesian Optimization is then employed to efficiently navigate this space, identifying the most effective model and hyperparameter combination for energy disaggregation.

\subsection{Bayesian Optimization for Optimal Model Selection and Hyperparameter Tuning}

One of the most significant challenges in implementing Machine Learning solutions lies in identifying the optimal model configuration. Traditionally, developers attempt to minimize the testing error by iteratively adjusting the model during development. A critical component of this process involves tuning \textit{hyperparameters}—predefined variables that determine the model’s structure and influence the training process. These hyperparameters must be set before training begins, and finding the best combination typically requires a careful and iterative search.

To address this, the literature identifies four primary hyperparameter optimization strategies: Manual Search, Grid Search, Random Search, and Bayesian Optimization. Among these, Bayesian Optimization has gained prominence for its efficiency and effectiveness. Unlike Grid or Random Search—which treat each evaluation independently—Bayesian methods utilize information from previous evaluations to inform future selections, making the search more intelligent and sample-efficient.

In this work, we adopt Bayesian Optimization specifically through the Hyperopt library for automated hyperparameter tuning. Hyperopt is an open-source Python framework for \textit{Distributed Asynchronous Hyperparameter Optimization}, employing \textit{Sequential Model-Based Optimization (SMBO)} with \textit{Tree-structured Parzen Estimators (TPE)}. It excels in handling a variety of variable types, search scales, and conditional dependencies in complex search spaces.

\begin{figure*}[h]
\begin{center}
    \centering
    \includegraphics[width=1\linewidth]{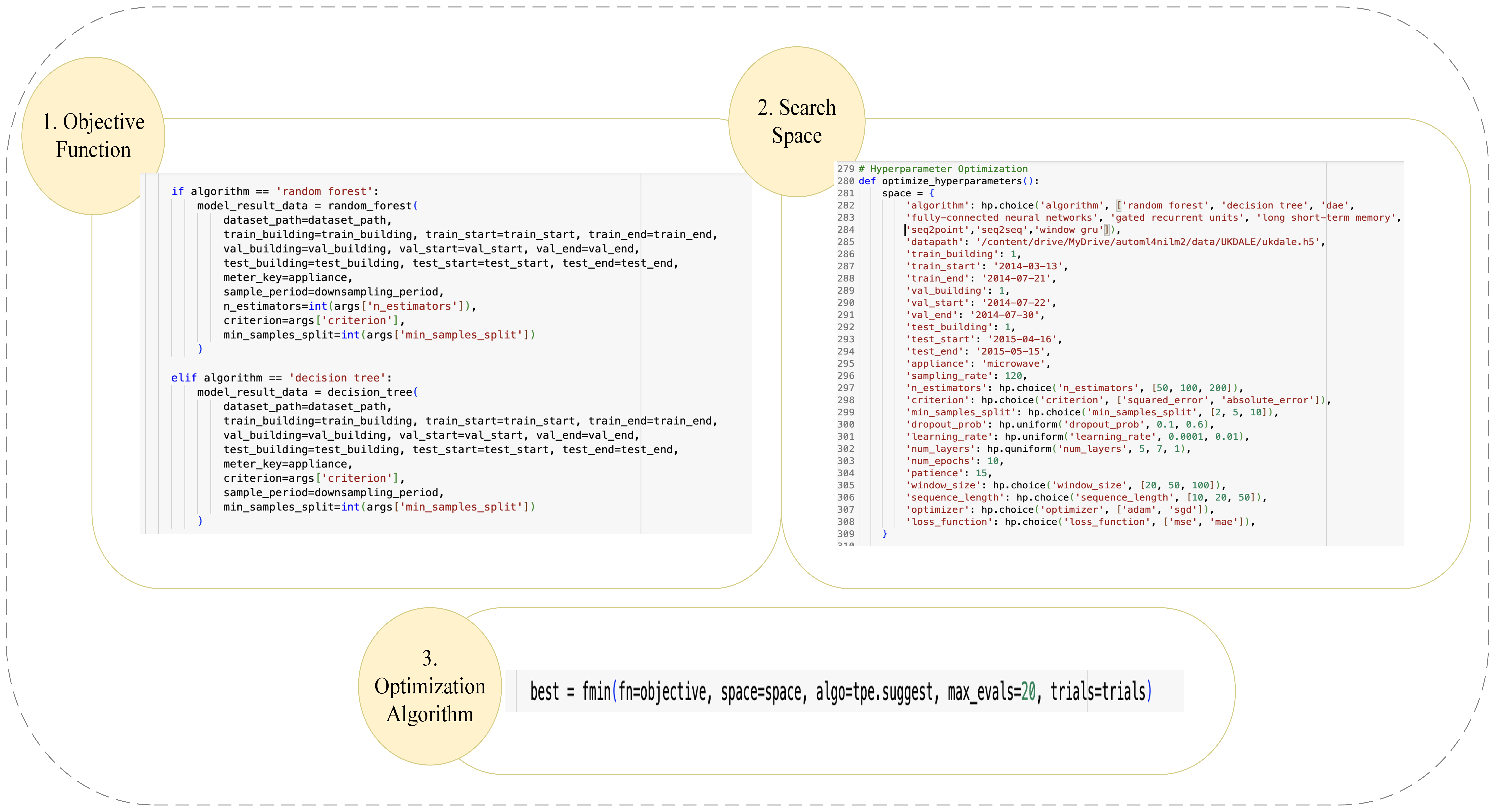}
    \caption{Three Main Components for Bayesian Optimization}
    \label{bo-components}
    \end{center}
\end{figure*}

Hyperopt's optimization process is defined by three main components, as shown in Figure \ref{bo-components}:
\begin{itemize}
    \item The search space, which specifies the allowable values for each hyperparameter;
    \item The objective function, which quantifies model performance (e.g., via accuracy or loss);
    \item The optimization algorithm, which intelligently navigates the search space using Bayesian principles.
Note that Figure \ref{bo-components} represents only a part of the overall project. To view the complete project, please refer to the AutoML4NILM GitHub repository.
\end{itemize}
In our study, we define a tree-structured search space and use TPE to optimize the \textit{Expected Improvement (EI)} criterion within the SMBO framework. The search spans a diverse set of ML models, including Decision Tree (DT), Random Forest (RF), Fully-Connected Neural Networks, Gated Recurrent Units (GRU), Window-based GRU, Seq2Seq, Seq2Point, Long Short-Term Memory networks (LSTM), Denoising Autoencoders (DAE), Factorial Hidden Markov Models (FHMM), and Combinatorial Optimization (CO).

In Section~\ref{experimental-results}, we provide detailed experimental results and benchmarking analyses across these models, demonstrating the effectiveness of our Bayesian AutoML approach for energy disaggregation.

\section{Experimental setup} \label{experimental-setup}
\textit{A. }\textit{Dataset}

Several energy disaggregation datasets have become available in recent years, marking a significant advancement in data-driven energy disaggregation. These datasets, collected from residential buildings across various countries, include sensor measurements such as active power, reactive power, current, and voltage. For NILM tasks, active power data is typically used. However, a key distinction among these datasets lies in their sampling frequencies. In this research, we utilized the UK-DALE dataset \cite{KellyKnottenbelt2015}, which is publicly available and specifically designed for studying disaggregation tasks. This dataset encompasses information on both the aggregate electricity demand of the buildings and the individual appliance-level ground truth demand. It consists of data from five buildings in the United Kingdom, spanning the years 2013 to 2015. The sampling rate for the whole-house measurements in this dataset is 16 kHz, while for individual appliances, it is 1/6 Hz, indicating a different sampling period. 

\textit{B. }\textit{Metrics}

For evaluating the performance of our models, two key metrics were employed: Mean Absolute Error (MAE) and Classification Accuracy. These metrics are integral to NILM systems, particularly within the context of smart grid applications for smart cities, which is the central focus of this research.
\subsubsection{Mean Absolute Error (MAE)}
The MAE is considered the most appropriate metric for assessing energy disaggregation models. It quantifies the average absolute deviation between the predicted and actual power consumption of appliances over a set of time intervals. For a particular appliance $i$ across $N$ time instances, the MAE is calculated as:

\[
\text{MAE}_i = \frac{1}{N} \sum_{n=1}^{N} \left|s_{in} - \hat{s}_{in}\right|
\]

where $s_{in}$ denotes the actual power consumption and $\hat{s}_{in}$ is the predicted value at time $n$.
\subsubsection{Classification Accuracy}
Classification Accuracy serves as an additional metric to evaluate the model’s ability to correctly classify the operational states of appliances, such as ON or OFF. This metric is essential for appliances like lights, where states are binary, but can also extend to appliances like dishwashers or microwave ovens, which may require more than two states to capture their operational behavior.

The Classification Accuracy is computed as follows:

\[
\text{Classification Accuracy} = \frac{TP + TN}{P + N}
\]

where $TP$ represents True Positives, $TN$ stands for True Negatives, and $P$ and $N$ are the total counts of positive and negative instances, respectively. This metric reflects how well the model can predict the actual state of appliances, which is critical for understanding the model's performance in real-world energy monitoring applications. 

\section{Experimental results}\label{experimental-results}
\begin{figure}
\begin{center}
    \centering
    \includegraphics[width=1\linewidth]{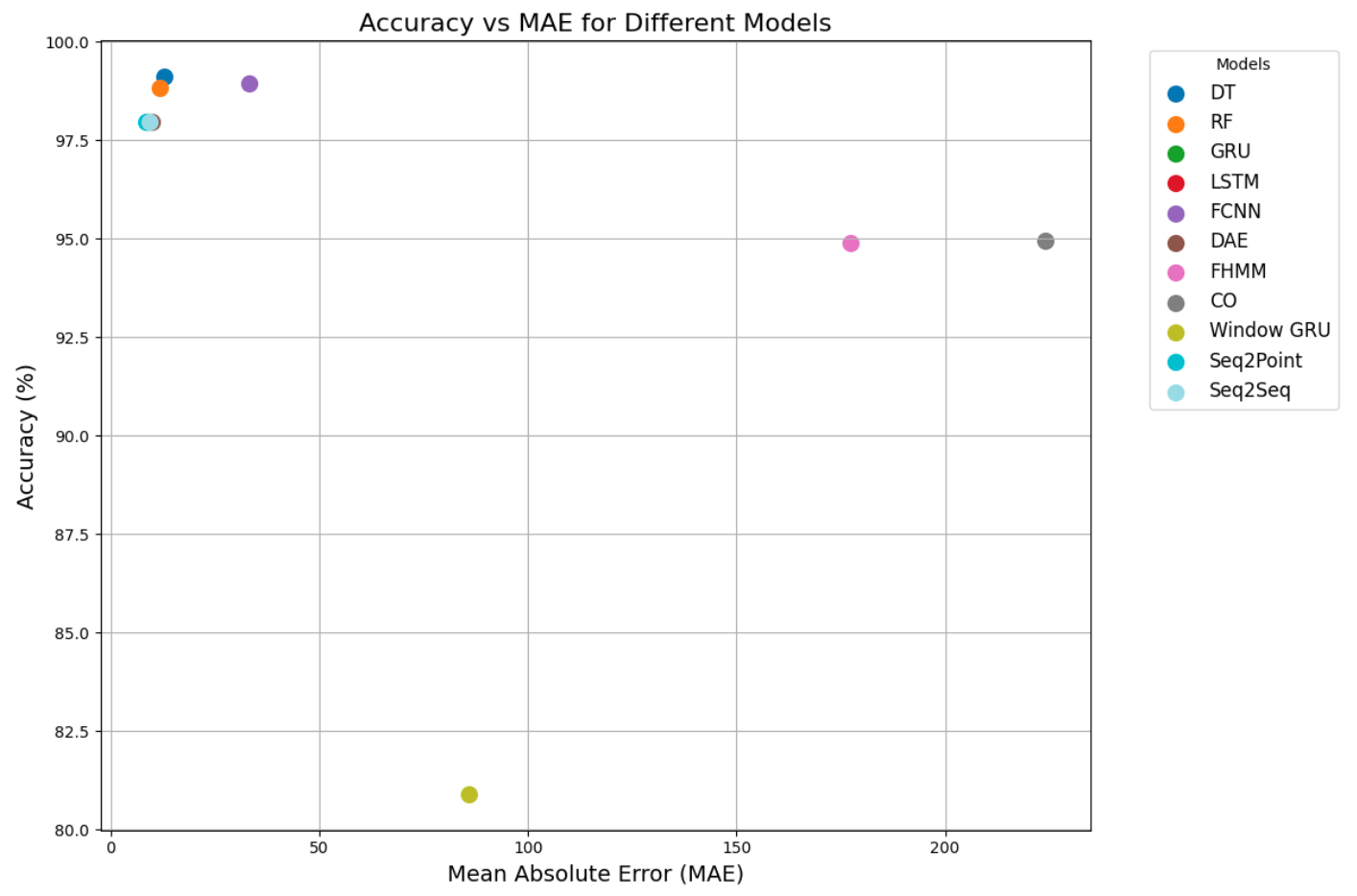}
    \caption{Classification Accuracy vs. MAE}
    \label{acc-mae-eachmodel}
    \end{center}
\end{figure}

\begin{figure}
\begin{center}
    \centering
    \includegraphics[width=1\linewidth]{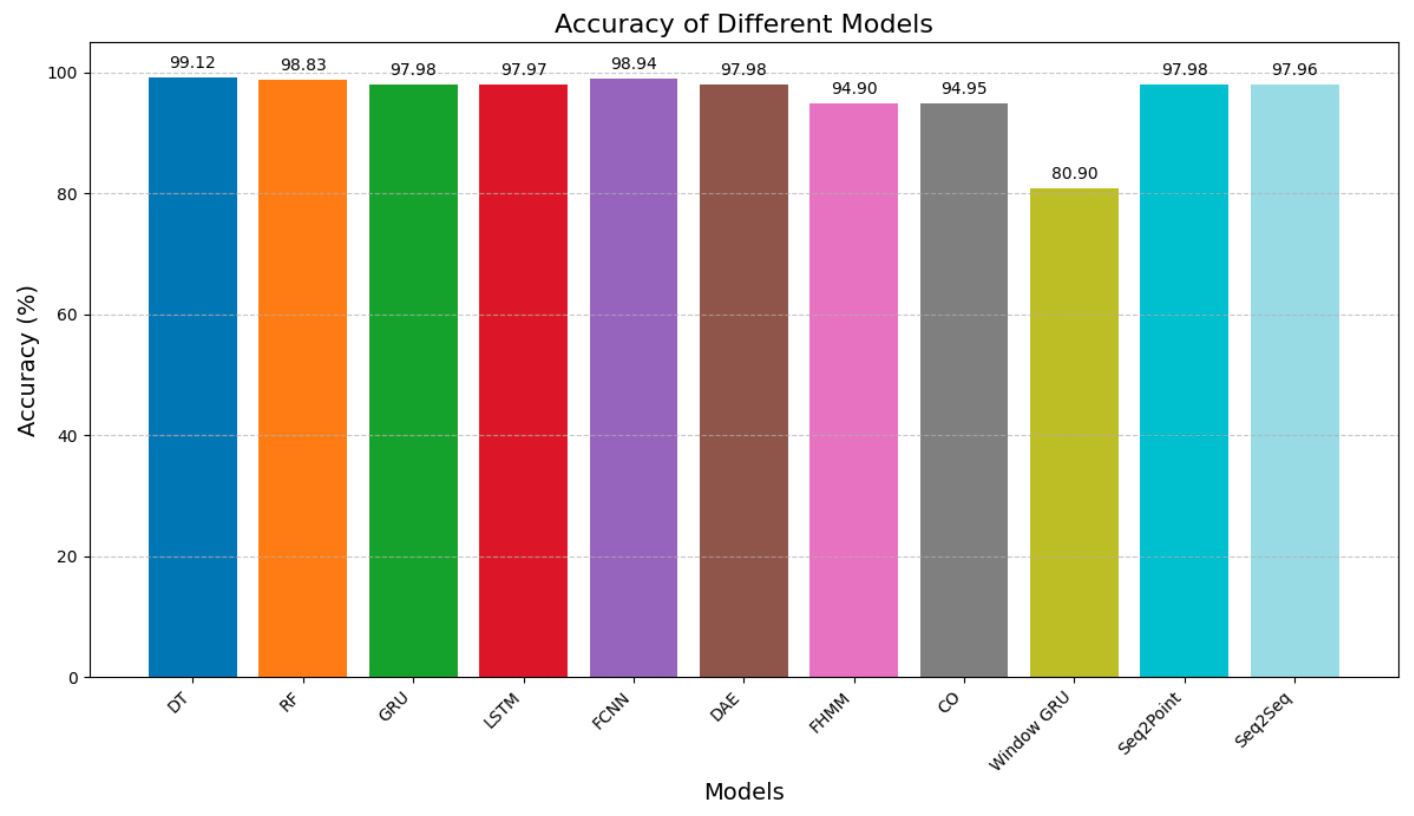}
    \caption{Classification Accuracy for each Model}
    \label{acc-eachmodel}
    \end{center}
\end{figure}

\begin{figure}
\begin{center}
    \centering
    \includegraphics[width=1\linewidth]{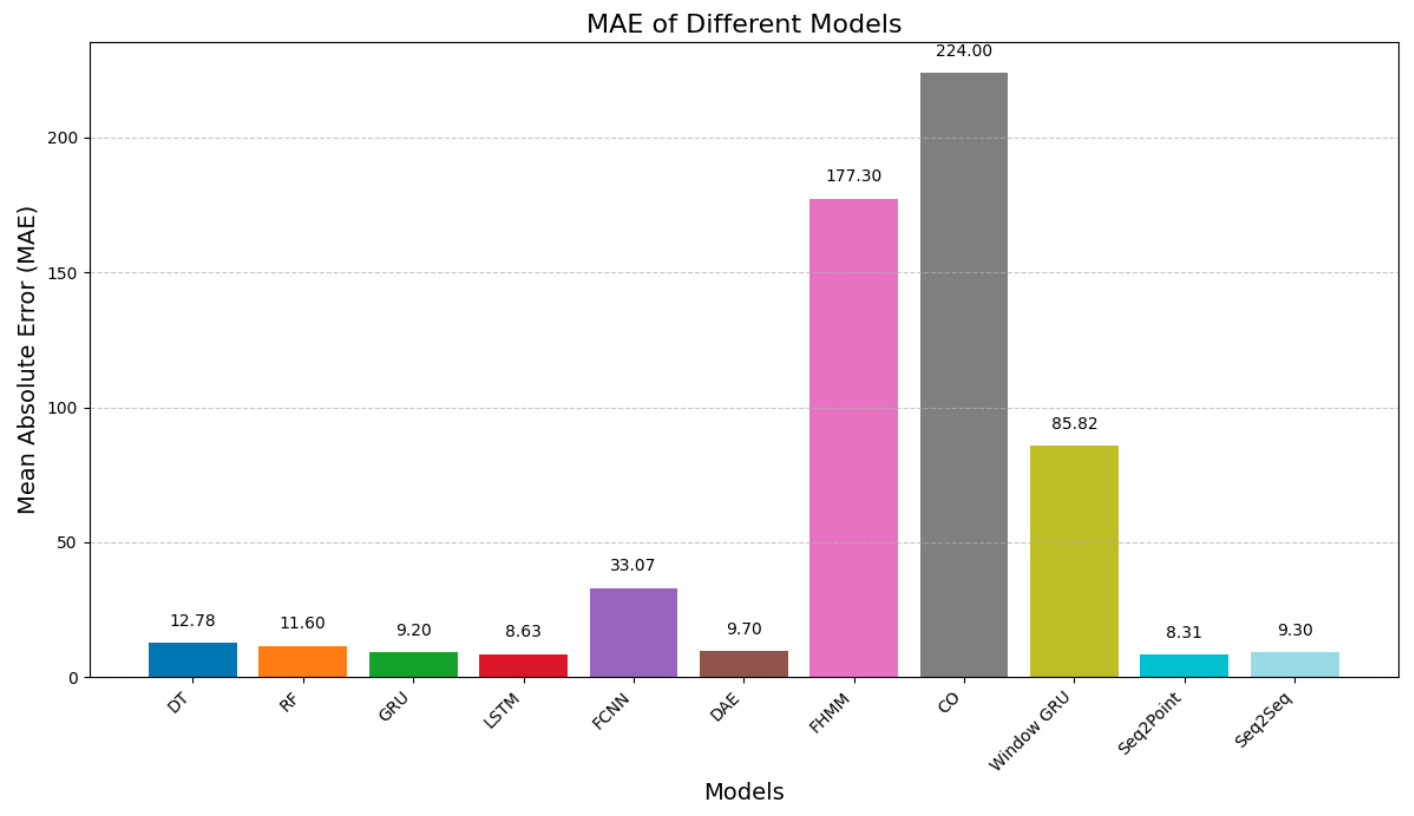}
    \caption{MAE for each Model}
    \label{mae-eachmodel}
    \end{center}
\end{figure}

As can be seen in Table \ref{Datasplit-table}, the data split for the UK-DALE dataset is as follows: the training set spans from March 13, 2014, to April 7, 2014 (25 days), the validation set covers April 7, 2014, to April 14, 2014 (7 days), and the test set starts from April 14, 2014, and goes until May 15, 2015, continuing for the remainder of the dataset. We use 20 epochs for model training. Additionally, a 60 Hz sampling period is chosen for disaggregation due to its balanced trade-off between capturing sufficient detail for appliance power consumption dynamics while reducing computational complexity, as faster sampling rates like 16 kHz would introduce unnecessary noise and higher computational cost. In Table \ref{each-alg-results}, the result of the accuracy and MAE of running each model separately is shown. Figure \ref{acc-mae-eachmodel} illustrates both the Classification Accuracy and the MAE for various approaches simultaneously. The diagram indicates that CO and, subsequently, FHMM perform the worst in terms of MAE. Additionally, the Window GRU model exhibits the poorest performance in terms of accuracy. Furthermore, Figures \ref{acc-eachmodel} and \ref{mae-eachmodel} provide individual visualizations of accuracy and MAE for each model when evaluated separately.

\begin{figure}
\begin{center}
    \centering
    \includegraphics[width=1\linewidth]{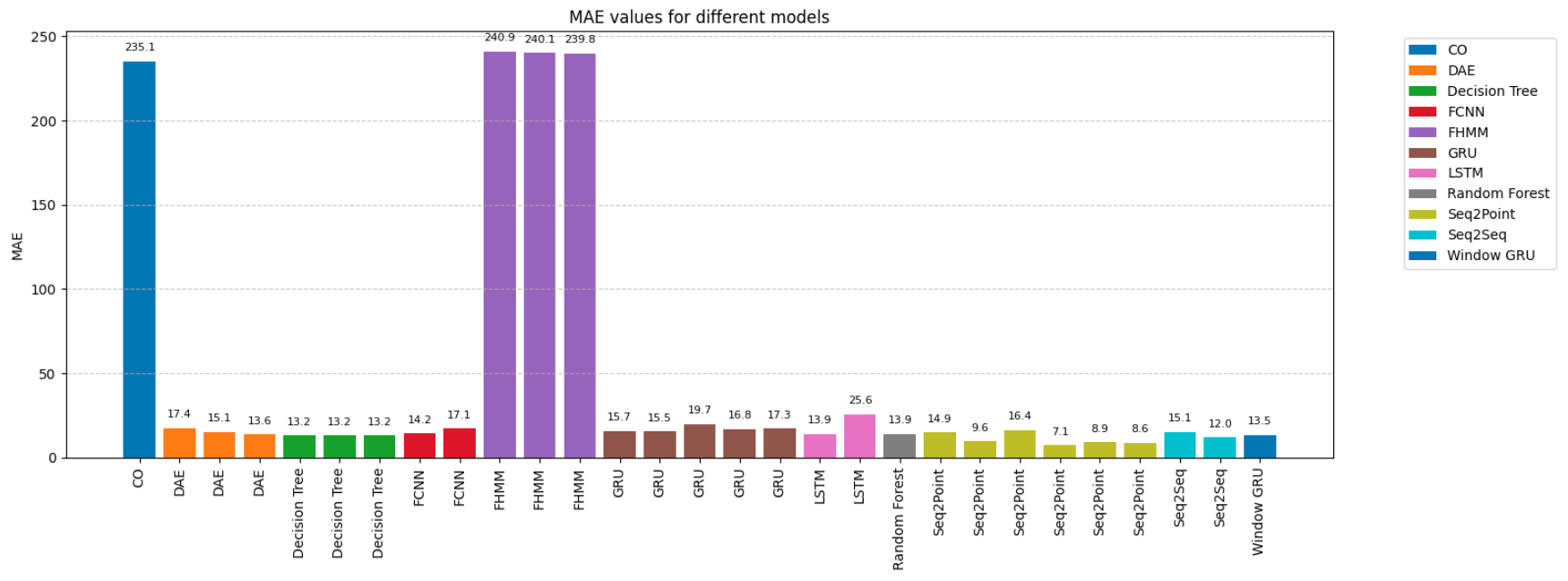}
    \caption{MAE for Each Model Executed in AutoML4NILM}
    \label{mae-automl4nilm-eachmodel}
    \end{center}
\end{figure}

\begin{table}

    \centering
    \begin{tabular}{|c|c|c|}
        \hline
        Data Split& Start Date& End Date\\ \hline
        Train& 2014-03-13& 2014-04-07 (25 days)\\ \hline
        Validation& 2014-04-07& 2014-04-14 (7 days)\\ \hline
        Test& 2014-04-14& 2015-05-15 (until the end)\\\hline
    \end{tabular}

    \caption{Data Splitting for UK-DALE dataset}
    
    \label{Datasplit-table}
    
\end{table}
\begin{table}[h!]
\centering
\begin{tabular}{|l|c|c|}
\hline
Model& Accuracy (\%)& MAE\\
\hline
Decision Tree& 99.12 & 12.78 \\
Random Forest& 98.83 & 11.6 \\
Gated Recurrent Units& 97.98 & 9.2 \\
Long Short-Term Memory& 97.97 & 8.63 \\
Fully-Connected Neural Networks& 98.94 & 33.07 \\
Denoising AutoEncoder & 97.98 & 9.7 \\
Factorial Hidden Markov Models& 94.9 & 177.3 \\
Combinatorial Optimization& 94.95 & 224 \\
Window GRU & 80.9 & 85.82 \\
Seq2Point & 97.98 & 8.31 \\
Seq2Seq & 97.96 & 9.3 \\
\hline
\end{tabular}
\caption{Performance comparison of different models based on Accuracy and MAE for NILM (Running Each Algorithm Individually )}
\label{each-alg-results}
\end{table}
The primary objective of AutoML4NILM was to identify the most effective machine learning model and its optimal hyperparameters, with a specific emphasis on minimizing the MAE. However, the framework also allows for the evaluation of other performance metrics, such as accuracy. The results shown in Table \ref{AutoML4NILM-each-alg-results} and also Figure \ref{mae-automl4nilm-eachmodel} indicate that the optimal model is Seq2Point, with the hyperparameters windowsize: 20, learningrate: 0.01, and optimizer: nadam, achieving the best performance in terms of MAE. The best MAE obtained was 7.12, representing the lowest MAE value among all models tested. These results are based on training for 20 epochs with a maximum of 30 evaluations during hyperparameter optimization.

It is important to note that due to limited computational resources, it was not feasible to conduct experiments with a larger number of evaluations, such as 100 max evaluations. Attempts to do so led to server connection crashes or PyCharm becoming unresponsive. Additionally, there are many other hyperparameter configurations that could potentially be tested. However, computational constraints also prevented extensive exploration; for instance, increasing the number of estimators in decision trees was not feasible, as it required more than 5 hours per epoch, ultimately causing the framework to fail during training.
\begin{table}[h!]
\centering
\begin{tabular}{|l|c|}
\hline
Model & MAE \\
\hline
Decision Tree & 13.17,\ 13.17,\ 13.17\\
Random Forest & 13.89 \\
Gated Recurrent Units & 15.70,\ 15.49,\ 19.69,\ 16.85,\ 17.27 \\
Long Short-Term Memory & 13.93,\ 25.60 \\
Fully-Connected Neural Networks & 14.19,\ 17.05 \\
Denoising AutoEncoder & 17.37,\ 15.14,\ 13.58 \\
Factorial Hidden Markov Models & 240.91,\ 240.15,\ 239.79 \\
Combinatorial Optimization & 235.07 \\
Window GRU & 13.51 \\
Seq2Point & 14.89,\ 9.59,\ 16.38,\ \textbf{7.12},\ 8.95,\ 8.59\\
Seq2Seq & 15.08,\ 12.02 \\
\hline
\end{tabular}
\caption{Results of AutoML4NILM for each algorithm with a maximum evaluation of 30 and 20 epochs}
\label{AutoML4NILM-each-alg-results}
\end{table}

\section{Conclusion and Future Works} \label{Conclusion}
In this study, we introduced AutoML4NILM, a versatile and extensible framework aimed at automating the processes of machine learning model selection and hyperparameter optimization for NILM applications. By incorporating reusable components, such as optimizers, models, datasets, and evaluation metrics, AutoML4NILM significantly streamlines the development of NILM models, minimizes manual effort, and facilitates more systematic experimentation.

Our methodology reflects principles of reuse-based software engineering and aligns with the full reuse model of software evolution and maintenance, where new capabilities are built by adapting and extending existing components. Through a focus on perfective maintenance, we continually enhance the framework by adding new algorithms, improving efficiency, and refining the hyperparameter search space.

The experimental results indicate that among the models assessed, the Seq2Point architecture achieved superior performance, exhibiting the lowest MAE across UK-DALE dataset. This finding underscores the efficacy of sequence-to-sequence models for precise energy disaggregation at the appliance level.

AutoML4NILM not only expedites research in the field of NILM but also establishes a structured reference architecture that meets the needs of both academic researchers and industry practitioners. Future directions will involve expanding the range of supported models, enhancing optimization strategies, and increasing adaptability to new NILM datasets and emerging challenges. Additionally, we plan to evaluate our approach using a broader set of metrics, as outlined in the GitHub repository for AutoML4NILM.
\section*{Data and Source Code Availability}
The source code is available in a GitHub repository \footnote{https://github.com/qas-lab/AutoML-for-NILM}. Also, in this research, we used the UK-DALE dataset \cite{KellyKnottenbelt2015}, which is publicly available and specifically designed for studying disaggregation tasks. 
\section{Acknowledgment}
In preparing this work, we used generative AI models and tools, including OpenAI models through ChatGPT, to assist with generating and revising content, including code and text. 
\bibliographystyle{ieeetr}
\bibliography{references}

\end{document}